\documentclass[12pt]{article}
 \usepackage{scicite}

 \usepackage{amssymb}
 \usepackage[pdftex]{graphicx}
\usepackage{times}
\newcommand\fpk{\mbox{$f_{\mathrm{pk}}$}}
\newcommand\ndyn{\mbox{$n^{\mathrm{dyn}}$}}
\newcommand\nidyn{\mbox{$n_{\mathrm I}^{\mathrm{dyn}}$}}
\newcommand\nstat{\mbox{$n^{\mathrm{stat}}$}}
\newcommand\nistat{\mbox{$n_{\mathrm I}^{\mathrm{stat}}$}}

\def\nuh{\mbox{$\nu_{\mathrm{H}}$}}
\def\nul{\mbox{$\nu_{\mathrm{L}}$}}

\def\nh{\mbox{$n_{\mathrm{H}}$}}
\def\nl{\mbox{$n_{\mathrm{L}}$}}
\def\ntot{\mbox{$n_{\mathrm{tot}}$}}

\def\to{t_{\mathrm{0}}}
\def\zo{Z_{\mathrm{0}}}

\def\sxx{\sigma_{\mathrm{xx}}}

\newenvironment{sciabstract}{%
\begin{quote} \bf}
{\end{quote}}

\title{ Wigner solid pinning modes tuned by fractional quantum Hall states of a nearby layer}

\author{A.\,T. Hatke$^1$, H.\,Deng$^2$,Yang\, Liu$^2$,  L.\,W. Engel$^{1\ast}$, \\
L.\,N. Pfeiffer$^2$, K.\,W. West$^2$, K.\,W. Baldwin$^2$, M. Shayegan$^2$ 
 \\
\normalsize
$^1${National High Magnetic Field Laboratory, Tallahassee, FL 32310, USA}\\
\normalsize$^2${Dept. of Electrical Engineering, Princeton University, Princeton, NJ 08544, USA}\\
 \\
\normalsize $^\ast$ to whom correspondence should be addressed, email:  engel@magnet.fsu.edu 
}
 \date{}
\begin{document}
\baselineskip24pt
  \maketitle

\begin{sciabstract}
We study a bilayer system hosting exotic many-body states of two-dimensional electron systems (2DESs) in close proximity but isolated from one another by a thin barrier. One 2DES has low electron density and forms a Wigner solid (WS) at high magnetic fields. The other has much higher density and, in the same field exhibits fractional quantum Hall states (FQHSs). The WS manifests microwave resonances which are understood as pinning modes, collective oscillations of the WS within the small but finite ubiquitous disorder. Our measurements reveal a striking evolution of the pinning mode frequencies of the WS layer with the formation of the FQHSs in the nearby layer, evincing a strong coupling between the WS pinning modes and the state of the 2DES in the adjacent layer, mediated by screening.

 %
\end{sciabstract}
 
 \newpage
 \subsection*{Introduction}
   Wigner solids occur when  electron-electron interaction dominates the zero-point or thermal motion of the carriers. They can  be accessed in extremely dilute systems in the absence of magnetic field or  in high magnetic field ($B$)  at sufficiently low Landau level filling,  $\nu$, at the termination of the FQHS series, where   Wigner solids have long been expected \cite{lozo,kunwc,rhim:2015}.  The magnetic-field-induced WS in a 2DES is of great interest and has been studied experimentally by a   variety of different techniques including pinning mode spectroscopy \cite{andrei:1988,williams:1991,li:2000,yewc,chen:2004,murthyrvw,wang:2012}, photoluminescence \cite{kukushkin:1994},  transport \cite{reentrant,goldman:1990,msreview,deng:2016}, NMR \cite{tiemann:2014} and time-dependent tunneling \cite{jang:2016}.      As a state stabilized by electron-electron interaction, it can be expected that a WS is strongly affected by nearby screening layers or its dielectric environment.    There are theoretical  works \cite{peeters:1984,spivak:2004}, concerning the  phase diagram of a 2DES in the presence of a    nearby metal gate, for  which the gate carries image charge that renders electron-electron interaction dipolar at distances exceeding the gate separation.  
For a WS near a higher-dielectric-constant  substrate, the screening is less strong, and image charge magnitude is less than  $|e|$, as was studied
\cite{peeters:1984,mistura:1997} for electrons separated from such substrates by  thin He films.
 
 \renewcommand{\baselinestretch}{1.2}
\begin{figure}[p]
\vspace{-0.2 in}
\includegraphics[width=\textwidth]{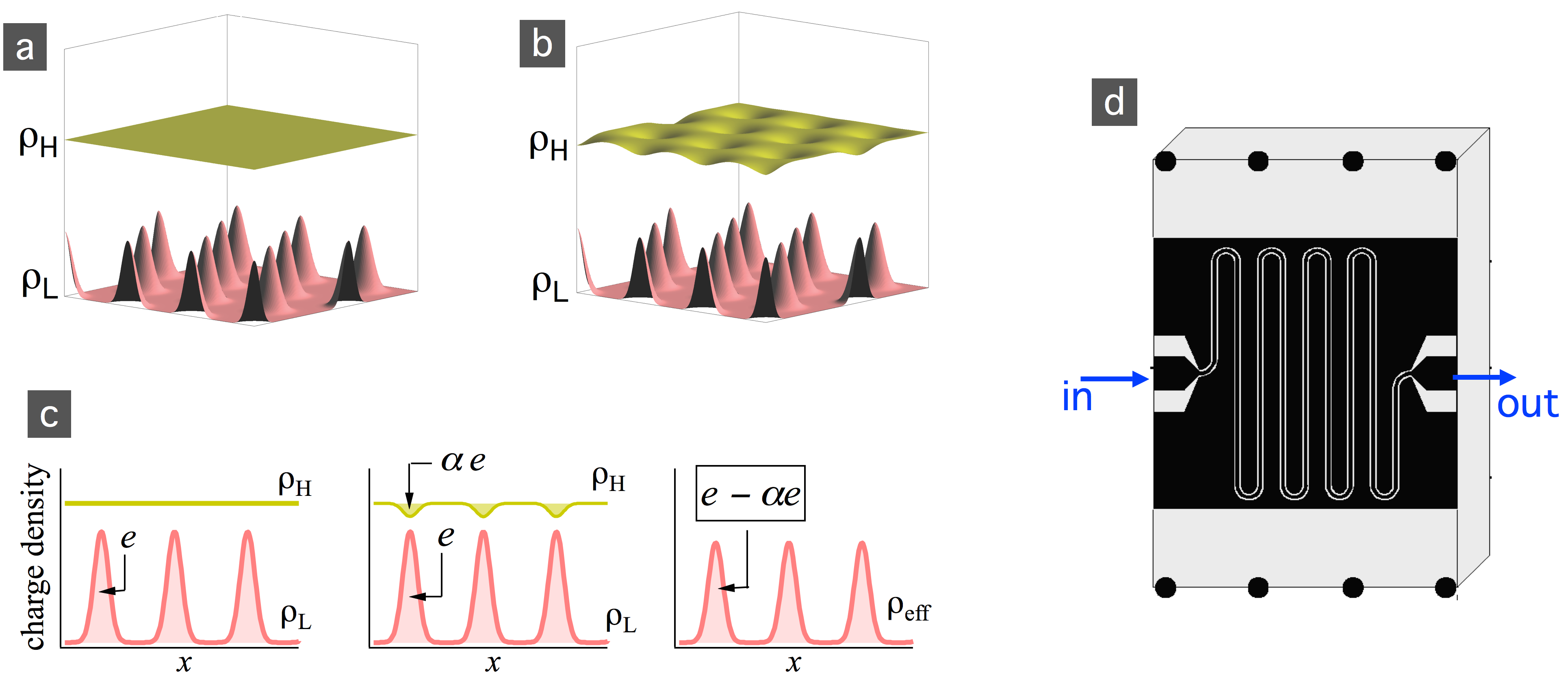}
\caption{  
  \textbf{Wigner solid (WS) close to composite fermion (CF) Fermi sea} 
 The bilayer system has a high-density (majority) top layer that hosts a CF Fermi sea  when its Landau filling \nuh\ is around 1/2, and exhibits FQHSs at odd-denominator fillings.   The low-density (minority) bottom layer has much smaller density compared to the majority layer and forms a   WS  when the majority layer is in the  regime of FQHSs. 
  Panels (a)-(c) show  local charge densities of the minority  layer and majority  layer,   $\rho_{\rm L}(x,y)$ and $\rho_{\rm H}(x,y)$.
 (a) The  charge densities without screening by the majority layer.  $\rho_{\rm L}$ shows the characteristic triangular Wigner lattice, but $\rho_{\rm H}$
 remains uniform, as in an incompressible liquid state.  
 (b)  Same as in  (a) but now the majority-layer density screens the WS and develops dimples, regions of locally reduced charge density, which act as opposite-signed  ``image'' charges.  
 (c) Three panels show cuts of figures l  (a) and (b), through a line of WS  electrons of charge $e$. The left panel is the incompressible-majority-layer situation as in (a). The middle panel shows the static dielectric response of a compressible majority layer to the WS of the minority layer.  The dimples, each with charge $-\alpha e$, develop in the majority layer, with $0\le\alpha\le1$.  The right panel illustrates our model of the screened WS: the charge of the image is modeled as summing with  the charge at its WS lattice site, creating an effective WS  charge per site of $(1-\alpha)e$.   
 (d) Schematic of the sample used for  microwave spectroscopy of WS pinning modes. The dark area is a metal-film coplanar waveguide (CPW) transmission line, through which microwaves are propagated. The CPW has  a driven center conductor and grounded side planes, and is capacitively coupled to the electrons in quantum wells.   A backgate   on the bottom of the sample   allows  the minority-layer density to be varied. \vspace*{-.7in}
}
\label{cartoon}
\end{figure}

Through pinning mode measurements  \cite{andrei:1988,williams:1991,li:2000,yewc,chen:2004,murthyrvw,wang:2012}, we   study here a 2D WS screened by a much larger density 2DES in a neighboring quantum well (QW). 
  Previous dc-transport studies \cite{deng:2016} of such density-asymmetric double wells have demonstrated the existence of a 
triangular-lattice WS in close proximity to a majority layer with  a composite fermion (CF) \cite{jainbook} metal near $\nu\sim 1/2$, by means of geometric resonance oscillations of the CFs acted on by   the WS.   Our work   considers the reverse,  and examines the effect of the CF metal and  nearby majority-layer FQHSs on the statics and pinning-mode dynamics of the WS.
In agreement with Ref. \cite{deng:2016} we  find strong pinning modes signifying the presence of a  WS.  The frequencies of these modes exhibit a remarkable dependence on the FQHSs formed in the nearby majority layer, allowing us to extract unique and unexpected information regarding the screening of the WS by this layer. 
The screening is closest to the dielectric-substrate case rather than the metal-gate case,  and can be modeled by image charges less than those at WS sites, as illustrated  in Fig. \ref{cartoon}.



%

%




\subsection*{Experimental Setup} 

Our samples contain two $30$-nm-wide GaAs QWs separated by a $10\,$-nm-thick, undoped  barrier layer of Al$_{0.24}$Ga$_{0.76}$As,   giving a center-to-center separation of $40\,$nm. 
The QWs are modulation-doped with Si $\delta$-layers asymmetrically: the bottom and top spacer layer thicknesses are $300\,$nm and $80\,$nm, respectively. 
This asymmetry leads to the different 2D electron densities in the QWs.  As cooled, the densities of  the top, high-density layer and the bottom, low-density layer are  $\nh \sim 15$ and  $\nl \sim 5.0$, in units of $10^{10}\,$cm$^{-2}$, which will be used for brevity in  the rest of the paper.  A bottom gate is used to control $\nl$.
As detailed in the Supplement, we obtained  \nh\ and $\nl$ following the procedure of Deng {\em et al.} \cite{deng:2016,deng:2017}, adapted for microwave conductivity measurements using the setup in Fig.  \ref{cartoon}(d).  $B$-dependent charge transfer between layers for samples like ours is possible, and occurs mainly for $\nuh>1$. To account for this, the total  density (\ntot), which does not change with $B$,  is obtained from low-$B$ Shubnikov-de Haas oscillations,   \nh\ comes from high-$B$ majority-layer FQHS positions, and $\nl$ in the $B$ range of interest is found by taking the difference between \ntot\ and \nh.

\begin{figure}[t]
\vspace{-0.4 in}
\begin{center} \includegraphics[width=.4\textwidth]{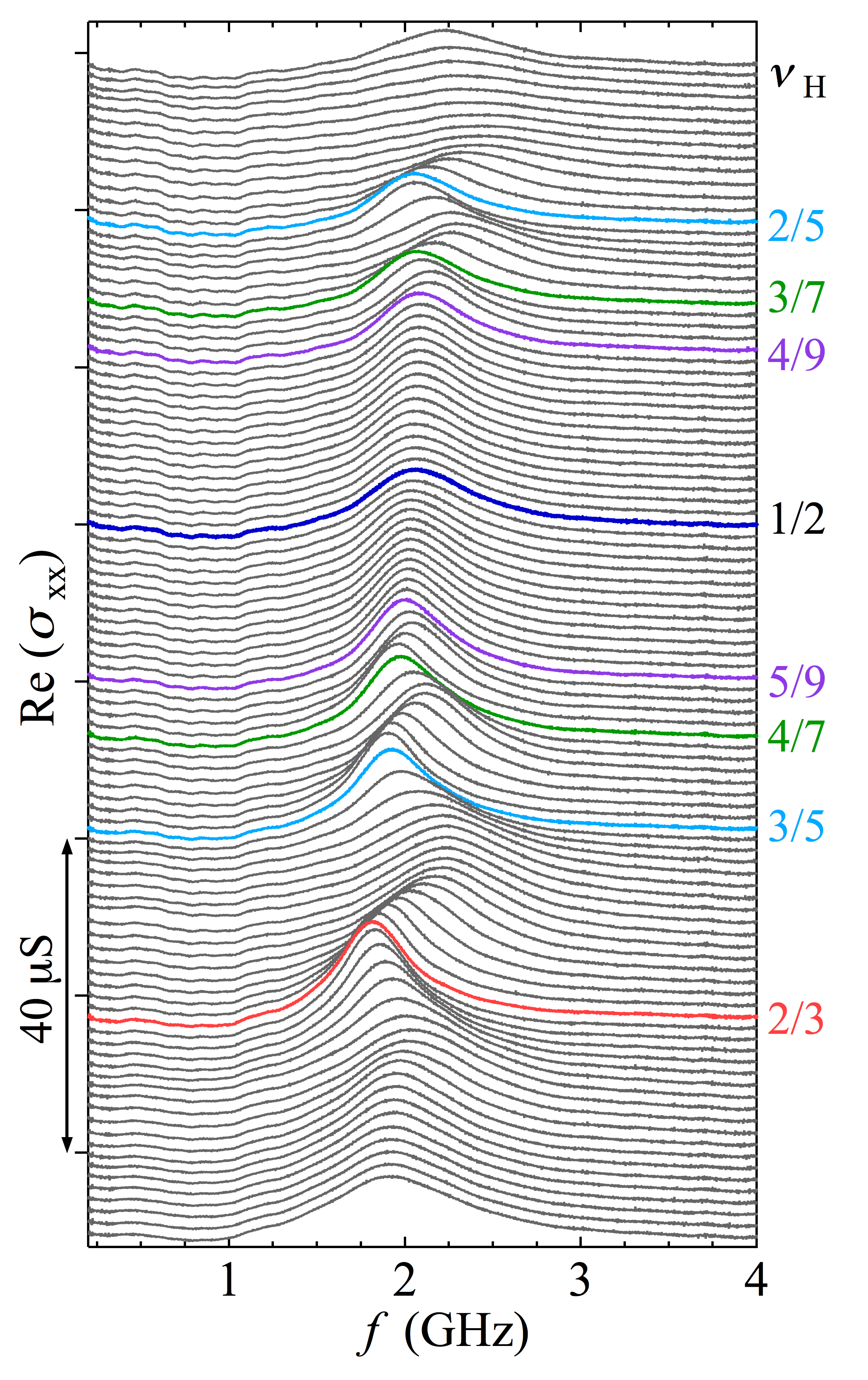}\end{center}
\vspace{-0.4in}
\caption{\textbf{Pinning mode spectra are strongly affected by majority layer.} 
Re\,$(\sxx)$ vs frequency, $f$, spectra  at many magnetic fields for majority and minority layer densities $\nh=15$ and $\nl=2.20$. Data were recorded in the low-power limit, and at the bath temperature of 50 mK.
Traces are vertically offset for clarity and  were taken in equal steps of $\nuh$ in the range $0.35 \le \nuh \le  0.75$ ($0.053 \le  \nul \le  0.113$).  The majority-layer filling $\nuh$ is labeled on the right axis.
}
\label{spectra}
\end{figure}

\begin{figure}[t]
\vspace{-0.5 in}
\begin{center}\includegraphics[width=.88\textwidth]{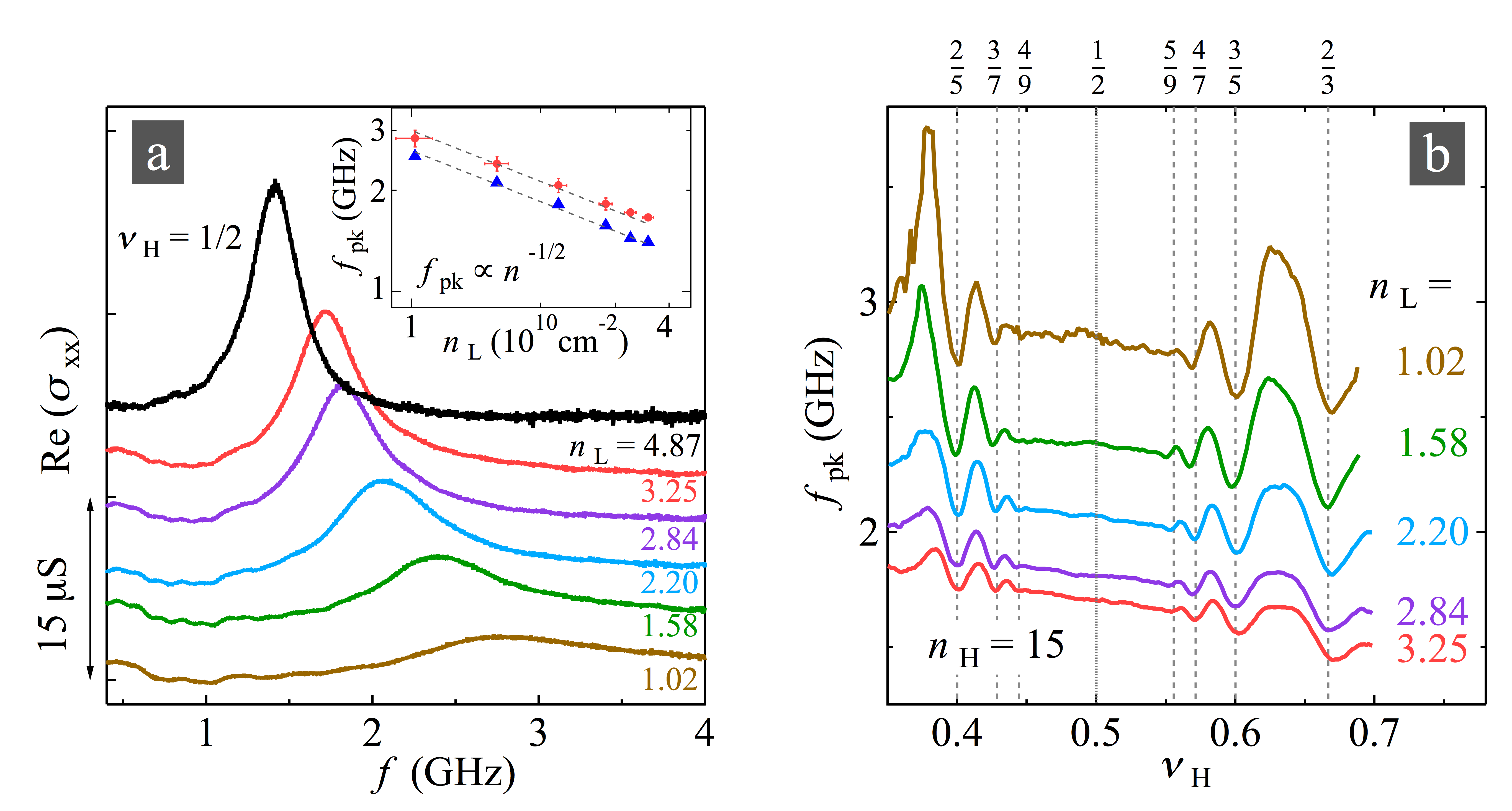}\end{center}
\vspace{-0.3in}
\caption{\textbf{ Pinning mode as minority-layer electron density, $\nl$, is varied. } 
(a)  Re\,$(\sxx)$ vs $ f $ spectra at fixed $\nuh=1/2$  ($\nh=15$)  at different $\nl$. Traces are offset vertically for clarity.  The density effect, typical for pinning modes, is evident as decreasing  \nl\ increases \fpk, while the  resonance amplitude decreases and the resonance broadens. Inset: extracted $\fpk$ vs $\nl$, on a log-log scale, at $\nuh=1/2$ (circles) and $2/3$ (triangles).  Dashed lines are fits   to  $\fpk\propto \nl^{-1/2}$.
 (b) $\fpk$ vs $\nuh$ at $\nh=15$ and different $\nl$; traces  {\em  not } vertically offset.
Vertical dashed lines mark  rational fractional fillings $\nuh$ of  FQHSs. The vertical dotted line marks 
 $\nuh=1/2$.   The upward overall increase of \fpk\ for each step down in  \nl\ is accompanied by the oscillations of \fpk\ vs \nuh, with minima at majority-layer FQHSs.  
}
\label{dfp}
\end{figure}  

\subsection*{Results}
The main result of this paper is contained in Fig. \,\ref{spectra}, which   shows  pinning modes exhibited by the  WS in the minority layer, as 
$B$ and hence the majority-layer filling, \nuh, are varied. 
The striking feature is   that, although the WS resides in the minority layer,  the pinning modes are clearly responding to the FQHSs   of the majority layer,
whose filling  $\nuh$ is marked at right in the figure.   
 The effect of the majority-layer state on the pinning modes makes it clear that screening of the WS is present. Throughout the measurement range the minority layer filling $\nul\le 0.113$,  well within the filling-factor range   of  WS  for  high-quality, 
single-layer 2DESs \cite{reentrant,goldman:1990,msreview,yewc,chen:2004}.
 
Figure \,\ref{dfp}(a)  illustrates  the  effect of  varying  $\nl$ on the pinning mode of the minority layer.   
  Re\,$[\sxx]$ vs $f$ spectra are shown at different $\nl$,  produced by changing backgate  voltage bias.   
Typical of pinning modes in a single-layer WS at low $\nu$ \cite{li:2000,yewc,wang:2012},  when $\nl$ decreases, the  peak frequency \fpk\ increases and  the resonance becomes broader and weaker.  We will refer to this behavior as the {\em density effect}. 
 Its explanation in weak-pinning theory \cite{chitra:2001,fertig:1999,fogler:2000} is that, as the WS softens at lower density, the carrier positions become more closely associated with disorder, and so on average experience a larger restoring force due to a small displacement.   The inset of Fig.\,\ref{dfp}(a)   shows the extracted $\fpk$ vs $\nl$.   The lines are fits to  $\fpk\propto n^{-1/2}$; such a  dependence 
  has been observed previously    \cite{li:2000,yewc,wang:2012} for  single-layer samples at low densities in the low-$\nu$ WS range. 
 
To highlight  the clear response of the pinning mode   to the majority-layer state,  most strikingly the reduction of   \fpk\ when a FQHS develops  in the majority layer at its odd-denominator fillings $\nuh= 2/5,3/7,4/7,3/5$ and $2/3$, in 
Fig. \,\ref{dfp}(b we show \fpk\ as a function of $\nuh$ for various $\nl$.
As $\nl$ decreases,  the overall $\fpk$ curves shift upward   over the entire $\nuh$   range.   The oscillation amplitudes of  \fpk\ seen in Fig.\,\ref{dfp}(b) at FQHSs of the majority layer  become more pronounced when \nl\  decreases.   
This is occurring  as the spacing of the minority-layer WS electrons  exceeds  the 40 nm interlayer  separation of the double-QW structure.
For example, at $\nl=3.25$ and $1.02$, the  triangular WS  lattice constant  is $a=60$ and $106$ nm, respectively. 

The  FQHS minima in Fig. \,\ref{dfp}(b) appear on top of  a weak decreasing background: for each trace, the \fpk\  oscillations,   and also   its featureless region between $\nuh=0.46$ and $0.54$,  are superimposed on a gradual decrease with $\nuh$.   
The decrease is similar for each trace, hence insensitive to $\nl$. 
In light of this insensitivity, we  ascribe the decreasing background to effects intrinsic to the minority layer. For example, such  effects could  be a change in the WS stiffness \cite{rhim:2015}   or a change in the disorder coupling \cite{chitra:2001,fertig:1999,fogler:2000}  due to a change in the magnetic length (size of the carrier).  Single-layer WSs are known  to show weak  dependence of \fpk\ on $B$ over wide ranges of Landau filling \cite{yewc}.

\begin{figure}[t]
\vspace{-0.2 in}
\begin{center}\hspace*{-.03in}\includegraphics[width=0.8\textwidth]{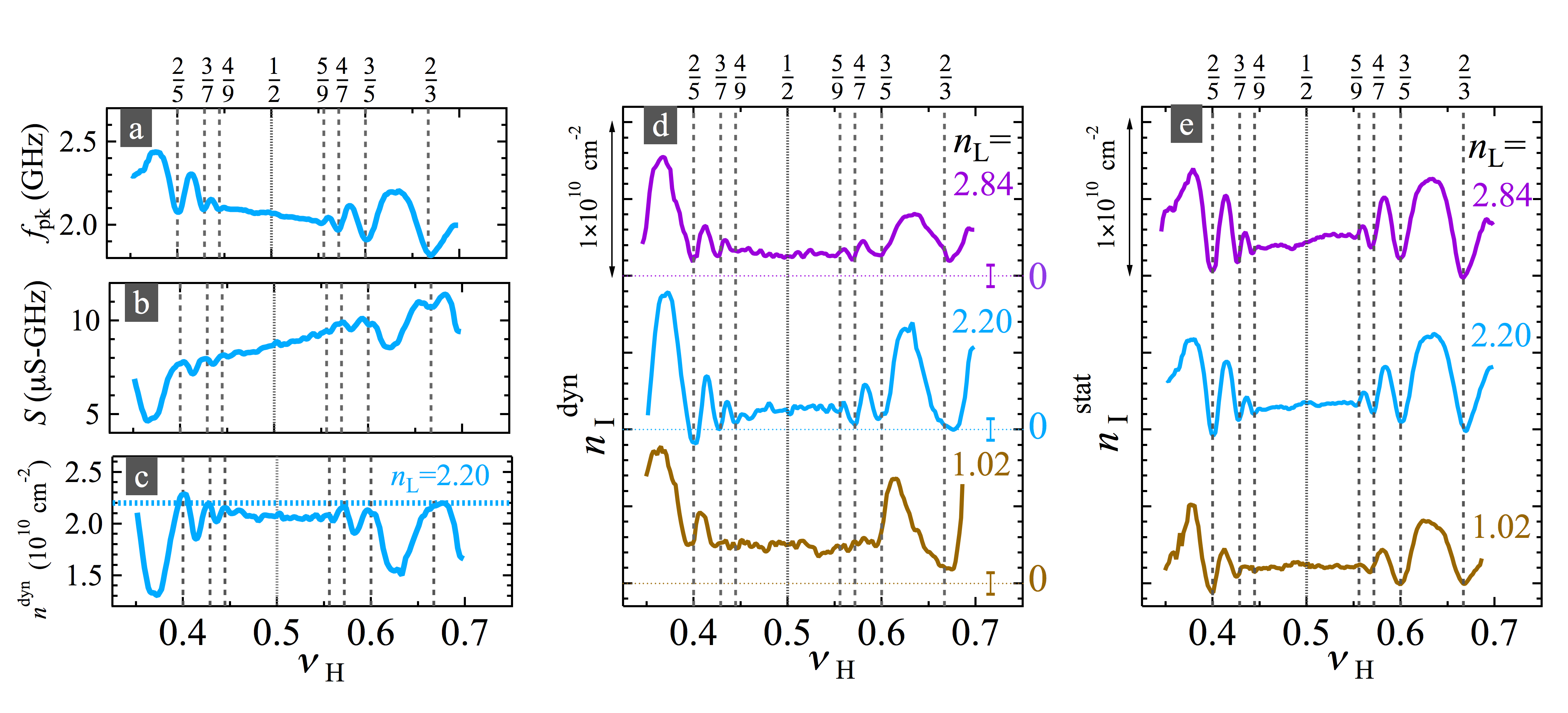}\end{center}
\vspace{-0.2 in}
\caption{{\bf Effective WS density and image charge densities.}
Panels (a)-(c) show  plots  of several quantities vs \nuh\ for $\nl=2.20$,  to illustrate  the determination of the dynamic effective WS density \ndyn:  (a)  shows \fpk, and   (b) shows $S$, the integrated Re$[\sxx]$ vs $f$, and  (c) shows \ndyn\ deduced from the pinning mode sum rule $\ndyn=(2B/\pi e) (S/\fpk)$. The overall downward or upward drifts respectively in \fpk\  and $S$  vs \nuh\   are removed in (c),  and a comparatively flat \ndyn\ vs \nuh\ is observed, in which the FQHSs appear as peaks.
(d) shows the density, $\nidyn= ( \ndyn-\nl )$  of the  image charge that moves as the pinning resonance is excited, for three \nl\ values, plotted vs \nuh.   The data are offset for clarity, and the respective zeroes of the traces are shown as lines with error bars.  
(e) shows the variation of the static image charge density, \nistat,   with \nuh. Traces are offset for clarity. 
\label{deltan}}
\end{figure}


\subsection*{Discussion}
 
Our  interpretation of the data  relies on the picture of Fig. \ref{cartoon}, in which, above the pinned WS lattice sites in the minority layer, the majority-layer local charge density   develops ``image" charge minima.  
  The amount of charge in each image depends on the {\em static} dielectric response of the majority layer, not its conductivity.  The ability of the image charge to follow the  WS site charge dynamically as the pinning mode is driven, on the other hand,  depends on the local conductivity of the majority layer as well. 
At each WS lattice site, there is then  a combination of an   image charge  with the corresponding charge in the WS.  This combined object has a dipole moment, but, because of the finite majority-layer local compressibility, it can also have a nonzero charge.  
We will characterize our pinning mode data in terms of charge densities.    \nstat\  denotes the static charge density of the combined  charges, and \ndyn\  denotes 
 the (dynamic) areal charge  density  that moves as the pinning mode is driven. 
 Like $\nl$, \nstat\ and \ndyn\ are given in units of  $10^{-10}$ cm$^{-2}$.  A static polarizability, $\alpha$, as in the caption of Fig.\,\ref{cartoon}, can be defined as   $\alpha=\nstat/\nl$. 
 
By means of the  pinning mode  sum rule \cite{fukuyama:1978}, $ \ndyn=   (2B/\pi e) (S/\fpk)$, where $S$ is the integrated Re\,$[\sxx]$ vs frequency, $f$, for the resonance.   Figures \ref{deltan}(a-c) show, for $\nl=2.20$,    how \ndyn\ is determined: \fpk\ vs \nuh\ in panel (a)  and  $S$ from panel (b)   produce \ndyn\ in panel (c) by use of the sum rule.   $S$ tends to  increase as  \fpk\ decreases and vice versa.  $S$ is increased near the majority-layer FQHS states, reflecting a lack  of available cancelling image charge at these low-compressibility states. In panel (c), near the peaks at the  
  most developed FQHSs ($\nuh=2/3$ and $2/5$),  \ndyn\ approaches \nl, which is  shown as a horizontal line.
  The difference of   \nl\  and \ndyn\ is the  image charge density in the majority layer that is moving along with the electrons of the WS, reducing the total current due to the resonance. We call $(\nl-\ndyn)$ the dynamic {\em image} charge density, \nidyn. It is graphed vs \nuh\  for $\nl=1.02,2.20$ and $2.84$  
  in Fig. \,\ref{deltan}(d).    
  \nidyn\ shows minima at the majority-layer FQHSs, reflecting their  small compressibility and small conductivity.

 The {\em static} image charge density \nistat, obtained as $(\nl-\nstat)$,   is of particular interest because of its sensitivity to the dielectric response of  the majority layer without the influence of the conductivity. It is plotted in Fig.\,\ref{deltan}(e).   While there is no
 direct method to measure \nstat\ or  \nistat, we can estimate  their  variations  as $\nuh$ sweeps through the FQHSs of the majority layer.   We obtain \nstat\   independently from \ndyn, from the \fpk\ data of Fig.\,\ref{dfp}(b) alone.   This is possible because in weak-pinning theories \cite{chitra:2001,fertig:1999,fogler:2000}, \fpk\   is solely determined by the    stiffness of the WS and the  disorder acting on it.  Increasing the  density of a WS  raises its stiffness.  As described in the Supplement,   the density-effect law,  $\fpk\propto \nl^{-1/2}$ is inverted  to find \nistat\ to within an additive constant.     By obtaining \nstat\ from the density-effect law we are treating \nstat\ as  if the image charge were on the same layer as the WS; because there is an interlayer separation  on the order of the WS lattice constant, this will overestimate the effect of \nistat, so that the \nistat\ we obtain are lower-limit estimates of the true image charge.  
 
 A valid low estimate for the absolute \nistat\ at $\nuh=1/2$, is obtained by neglecting the finite compressibility of the majority layer at $\nu=2/3$ and taking 
 $\nistat(\nuh\,=\,2/3)=0$.   For the three \nl\ values of Fig.\,\ref{deltan}, 1.02, 2.20 and 2.84, we find this 
 low-estimate $\nistat(\nuh\,=\,1/2)$ is about 10\% of \nl.  The values of $\nidyn(\nuh=1/2)$  are on the order of their error, also about 10\% of  \nl. Overall, we find the variations of \nistat\ and \nidyn\ to be of similar size for most \nuh.  This implies that the image charge in the majority layer  moves with   
the WS   as the resonance is driven.


 In summary we study a WS separated from  FQHSs by a distance comparable to its lattice constant.    
 We observe a pinning mode from the  minority-layer WS,  indicating its existence even in the presence of the nearby, screening majority layer.   
The pinning mode is strongly affected by  the majority-layer FQHSs, exhibiting  a reduction in \fpk\ with an increase in $S$ around FQHEs. We find that these phenomena can be  modeled by considering image charges in the majority layer, and regarding them as reducing the WS charge.   The results indicate that in large part the image charge oscillates as the pinning mode is driven. The image charge is assessed to be about 10\% of the WS charge near $\nuh=1/2$, but substantially larger elsewhere, particularly at the transitions between FQHSs.

%

\subsection*{Methods}

We performed  microwave spectroscopy \cite{li:2000,yewc,chen:2004,murthyrvw,wang:2012} using  a coplanar waveguide (CPW) patterned in Cr:Au film on the top surface of the sample.
A top view schematic of the measurement is shown in Fig.\,\ref{cartoon}\,(d).
We calculate the diagonal conductivity as $ \sigma_{xx} (f) = (s/ l \zo) \ln (t/\to)$, where $s=30\ \mu$m is the distance between the center conductor and ground plane, $l=28\,$mm is the length of the CPW, $\zo=50\,\Omega$ is the characteristic impedance without the 2DES, $t$ is the transmitted signal amplitude and $\to$ is the normalizing amplitude.   
The  microwave measurements were carried out in the low-power limit, such that the results are  not sensitive to the excitation power  at our bath temperature of $T=50\,$mK.  

 \subsection*{Acknowledgements}
 
We thank Ju-Hyun Park and Glover Jones for their expert technical assistance, and J. P. Eisenstein  for  discussions. 
The microwave spectroscopy work at the National High Magnetic Field Laboratory (NHMFL) was supported through Department of Energy Basic Energy Sciences (DOE-BES) grant DE-FG02-05-ER46212 at NHMFL/FSU.   
The  NHMFL is supported by National Science Foundation (NSF) Cooperative Agreement Nos. DMR-1157490 and  DMR-1644779, by the State of Florida, and by the DOE.  The work at Princeton University was funded by the Gordon and Betty Moore Foundation through the EPiQS initiative Grant GBMF4420, and by the DOE BES grant DE-FG02-00-ER45841 and the NSF through grant DMR-1709076 and MRSEC Grant DMR-1420541. Data displayed in this manuscript will be available by email request to engel@magnet.fsu.edu.

\subsection*{Contributions}
A.T.H. conceived and designed the experiment, performed the microwave measurements, analyzed the data and co-wrote the manuscript. L.W.E. conceived and designed the experiment, discussed data analysis and co-wrote the manuscript.   H.D., Y.L.  and M. S. conceived the experiment, discussed data analysis  and co-wrote the manuscript.   L.N.P., K.W.W. and K.W.B. were responsible for the growth of the samples.  
\bibliography{bib_ah}
\bibliographystyle{ScienceAdvances}

\end{document}


\baselineskip24pt
  \maketitle
\subsection*{\bf Obtaining \nh\ and \nl}

\begin{figure}[h]
\hfil\includegraphics[width=3in]{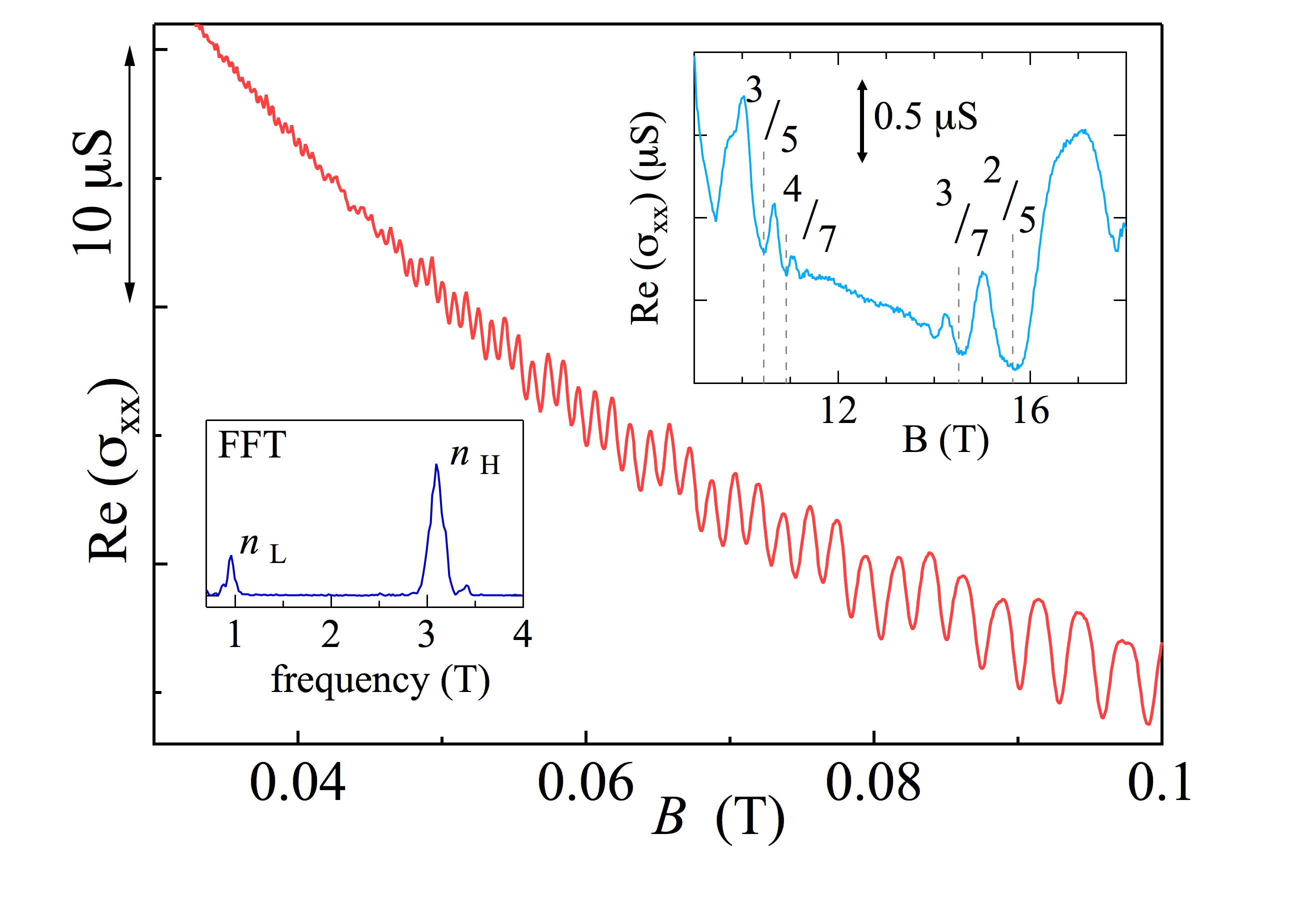}\hfil
\caption{
  {\bf Data  used to determine electron densities  of the double quantum well.}  
  Real part of the conductivity, Re\,$(\sxx)$, vs $B$ at fixed microwave frequency, $f=500\,$MHz, which exhibits two oscillation frequencies that correspond to Shubnikov-de Haas  oscillations arising from the two wells.
The upper inset presents   Re\,$(\sxx)$ at high $B$, exhibiting  the majority-layer FQHSs.  The lower inset shows the fast Fourier transform (FFT)  of the   Re\,$(\sxx)$ vs $1/B$ oscillations. 
}
\label{suppsdh}
\end{figure}

We obtained  \nh\ and $\nl$ following the procedure of Deng {\em et al.} \cite{deng:2016,deng:2017}, adapted for microwave conductivity measurements.  
For each backgate bias we measured   Re\,$(\sxx)$ vs magnetic field, $B$, at a constant frequency, $f=500\,$MHz,  for  $0.04\leq B\leq0.15\,$T.    Figure \ref{suppsdh} shows a representative trace. 
The magnetoconductivity exhibits Shubnikov-de Haas oscillations   with a fast oscillation corresponding to the majority layer superimposed on a slower oscillation corresponding to the minority layer.
The   frequencies of the two distinct peaks obtained from a fast Fourier transform, like that shown in the lower inset of
 Fig. \ref{suppsdh}, directly yield  $ \nh$ and $\nl $, whose  sum gives the total density, $\ntot$. 

In double quantum wells with unequal densities,   increasing $B$ can cause a charge transfer between the layers,  particularly until  $B$ is large  enough that   both wells are in the lowest Landau level \cite{deng:2016,deng:2017,eisenstein:1994} .
We obtain $\nh$ at sufficiently high $B$ for $\nuh<1$ through the magnetic-field positions of the well-defined FQHS minima in  magnetoconductivity, Re\,$[\sxx(B)]$,  from the majority layer, as shown in the upper inset of Fig.  \ref{suppsdh}.  The positions of these minima are consistent with a constant high-magnetic-field  \nh,  indicating to within  error of about 1\% that there is no charge transfer at high $B$. 
Assuming $\ntot$ is independent of $B$, by subtracting the extracted high-field  $\nh$ from $\ntot$, we  obtain  $\nl$ in the magnetic field range of interest.

\subsection*{ The possibility of interlayer charge transfer changing the minority-layer density due to the majority-layer FQHSs }
 
The changes in WS density  required to produce dips in \fpk\ as we observed are only on the order of a percent or so of 
the majority layer density, but are of the wrong sign to be explained by static charge transfer driven by the FQHS gap energy, as considered in Refs.  \cite{deng:2017} and \cite{eisenstein:1994}.  Considering a charge transfer driven by a change of $B$, the  total density change  cannot change, so $\delta \nh  +\delta \nl=0$.   The change in voltage between the layers due to the transfer is $(e^2/C)\delta \nl=\delta \mu_{\rm L} - \delta \mu_ {\rm H}$, where $C$ is the geometric interlayer capacitance per unit area, $C= \epsilon\epsilon_0/d$, with  $d$  the layer separation. On combining the expressions we find the constant $e^2/C=\delta  \mu_{\rm L}/\delta \nl + \delta \mu_{\rm H}/\delta \nh$.

For a WS, $\partial  \mu_{\rm L}/\partial\nl $ is expected to be negative, but to increase with \nl.  A classical WS \cite{fano:1988,eisenstein:1994} would have  $\partial \mu_{\rm L}/\partial\nl \propto -\nl^{-1/2} $. 
At a majority-layer FQHS, $\partial \mu_{\rm H}/\partial \nh$ exhibits a maximum \cite{eisenstein:1994}, so that a charge transfer would produce a minimum in  $\partial  \mu_{\rm L}/\partial\nl $.   But in a WS, $\partial  \mu_{\rm L}/\partial\nl $ increases with \nl,  so the charge transfer would require a minimum in \nl.  This is contrary to our pinning mode data, which imply a maximum in 
effective WS charge density at a majority-layer FQHS.

\begin{figure}[t]
\vspace{-0.2 in}
\hfil\includegraphics[width=2.5in]{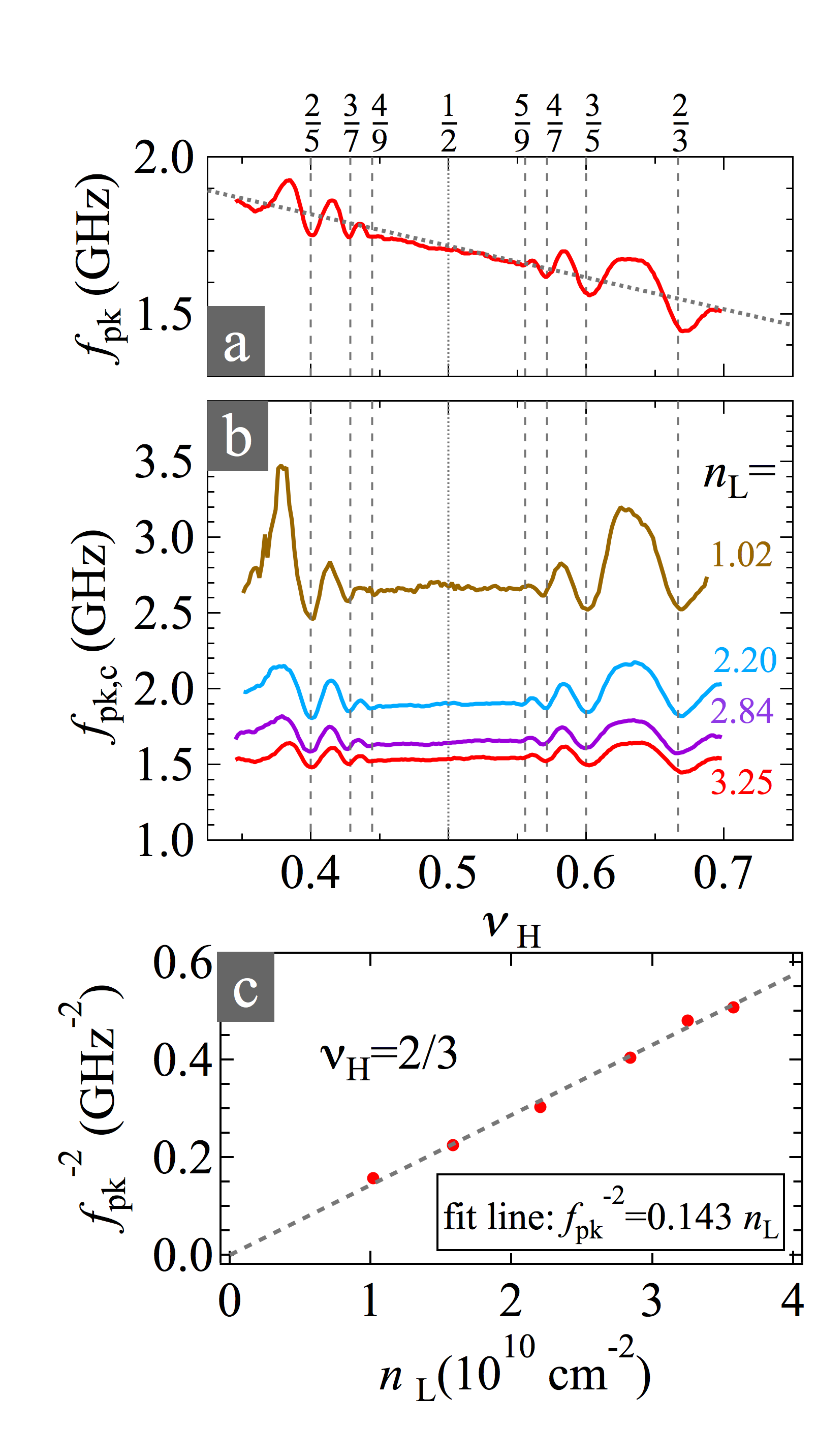}\hfil
 \vspace{-0.25 in}
\caption{\textbf{  Fits used to  obtain the static image charge density, \nistat.   } (a) \  \fpk vs \nuh, for $\nl=3.25$,  as in Figure 3(b) of the main text.  The dotted line is a fit to all data, and its  slope is used to subtract off the falling background from \fpk\ vs \nuh\ traces.  (b)   The resulting background-subtracted \fpk, which we call $f_{\rm pk,c}$ for  $\nl=1.02,2.20,2.84$ and $3.25$.  (c) The density-effect law,   $\fpk^{-2}=\beta \nl$, for the full range of \nl\ studied. \fpk\ is taken at $\nuh=2/3$.  $\beta$ from the  fit in (c) is used to  compute  $\nstat(\nuh)= \beta^{-1}f_{\rm pk,c}^{-2}(\nuh) $. 
\label{suppfits}
}
 \end{figure} 
 
\subsection*{\bf Determination of \nstat\ and \nistat\ from \fpk\ vs \nuh }
 
 We attributed the falling background in Fig. 3(b) of the main text to effects other than screening by the majority layer, so the first task is to remove this background empirically.   This is done by fitting the largest \nl\  ($\nl=3.25$) curve in Fig.\,3(b) to a line, as shown in Fig.\,\ref{suppfits}(a).  $\nl=3.25$ is chosen because its variations with \nuh\ are the smallest.    A line  with the slope of the fit in Fig.\,\ref{suppfits}(a) , but set to zero at $\nuh=2/3$, is  subtracted from the \fpk\  vs \nuh\ data
 to remove the falling background, {\em i.e.} to produce \fpk\ with the background subtracted,    $f_{\rm pk,c}$, shown in Fig.  \ref{suppfits}(b).  
 
 We then invert the density effect law to $\nstat=\beta f_{\rm pk,c}^{-2}$ to compute \nstat\ from $f_{\rm pk,c}$, as $\beta^{-1}f_{\rm pk,c}^{-2}(\nuh)$.  
  $\beta$ is found from the fit, shown in Fig. \ref{suppfits}(c), to  $\nl=\beta\fpk^{-2}$ over the full range of $\nl$ at $\nuh=2/3$ .   Because $\beta$ is found at $\nuh=2/3$, the \nstat\ computed at that filling is fixed to the fit line, which is close to \nl, so that  at $\nuh=2/3$  \nstat\ is arbitrarily set to near \nl\, and $\nistat=(\nstat-\nl)$ is  set to zero.

\bibliography{bib_ah}
\bibliographystyle{ScienceAdvances}